\pdfoutput=1

\documentclass[11pt]{article}

\usepackage{ACL2023}

\usepackage{times}
\usepackage{latexsym}
\usepackage{bigstrut}
\usepackage[hang,flushmargin,multiple]{footmisc}
\usepackage{booktabs} 

\usepackage[T1]{fontenc}

\usepackage[utf8]{inputenc}

\usepackage{microtype}

\usepackage{paralist}
\usepackage{caption}
\usepackage{subcaption}

\usepackage{comment}
\usepackage[flushleft]{threeparttable}
\usepackage{graphicx}
\usepackage{amsmath}
\usepackage{tabularx}
\usepackage{mathrsfs}
\usepackage{multirow}
\usepackage{algorithm}
\usepackage{enumitem}
\usepackage{xcolor}
\usepackage{listings}
\usepackage{hyperref}
\usepackage{xspace}
\usepackage{adjustbox}
\usepackage[most]{tcolorbox}
\usepackage{color, colortbl}

\usepackage{arydshln}
\newcommand{\eg}{\emph{e.g.,}~}

\newcommand{\tool}{\textsc{Avatar}\xspace}

\definecolor{dark-gray}{gray}{0.85}
\definecolor{light-gray}{gray}{0.95}
\definecolor{mygreen}{rgb}{0,0.4,0}
\definecolor{mygray}{rgb}{0.5,0.5,0.5}
\definecolor{mymauve}{rgb}{0.58,0,0.82}
\definecolor{myred}{rgb}{0.82, 0.1, 0.26}
\definecolor{whitesmoke}{rgb}{0.99, 0.99, 0.99}

\lstdefinestyle{CustomPy}{
    escapeinside={(*@}{@*)},
    belowcaptionskip=1\baselineskip,
    xleftmargin=1pt,
    xrightmargin=1pt,
    language=Python,
    numbersep=5pt,
    tabsize=4,
    showstringspaces=false,
    basicstyle=\small\ttfamily, 
    keywordstyle=\bf\color{mygreen},
    commentstyle=\color{purple},
    stringstyle=\color{red},
    identifierstyle=\color{black},
    numberstyle=\tiny\color{mygray},
    emph={int,char,double,float,unsigned,void,bool,boolean},
    emphstyle={\bf\color{myred}},
    emph=[2]{and, in,},
    emphstyle=[2]{\bf\color{violet}},
    emph=[3]{sortedCount, sorted_count},
    emphstyle=[3]{\bf\color{blue}},
    numbers=left,
    stepnumber=1,
    breaklines=true,
    backgroundcolor=\color{whitesmoke},
    literate={\ \ }{{\ }}1,
}

\lstdefinestyle{CustomJava}{
    belowcaptionskip=1\baselineskip,
    xleftmargin=1pt,
    xrightmargin=3pt,
    language=Java,
    numbersep=5pt,
    tabsize=4,
    showstringspaces=false,
    basicstyle=\small\ttfamily, 
    keywordstyle=\bf\color{mygreen},
    commentstyle=\color{purple},
    stringstyle=\color{red},
    identifierstyle=\color{black},
    numberstyle=\tiny\color{mygray},
    stringstyle=\color{mymauve},
    emph={int,char,double,float,unsigned,void,bool,boolean},
    emphstyle={\bf\color{myred}},
    emph=[2]{and, in,},
    emphstyle=[2]{\bf\color{violet}},
    emph=[3]{sortedCount, sorted_count},
    emphstyle=[3]{\bf\color{blue}},
    numbers=left,
    stepnumber=1,
    breaklines=true,
    backgroundcolor=\color{whitesmoke},
    literate={\ \ }{{\ }}1,
}

\makeatletter
\let\old@lstKV@SwitchCases\lstKV@SwitchCases
\def\lstKV@SwitchCases#1#2#3{}
\makeatother
\usepackage{lstlinebgrd}
\makeatletter
\let\lstKV@SwitchCases\old@lstKV@SwitchCases

\lst@Key{numbers}{none}{%
    \def\lst@PlaceNumber{\lst@linebgrd}%
    \lstKV@SwitchCases{#1}%
    {none:\\%
     left:\def\lst@PlaceNumber{\llap{\normalfont
                \lst@numberstyle{\thelstnumber}\kern\lst@numbersep}\lst@linebgrd}\\%
     right:\def\lst@PlaceNumber{\rlap{\normalfont
                \kern\linewidth \kern\lst@numbersep
                \lst@numberstyle{\thelstnumber}}\lst@linebgrd}%
    }{\PackageError{Listings}{Numbers #1 unknown}\@ehc}}
\makeatother

\newcount\myloopcounter

\newcommand{\repeatit}[2][10]{%
  \myloopcounter0
  \loop\ifnum\myloopcounter < #1 
  #2%
  \advance\myloopcounter by 1 %
  \repeat 
}

\usepackage{titlesec}
\titlespacing{\paragraph}{%
  0pt}{
  0.2\baselineskip}{
  1em}%

\title{\tool: A Parallel Corpus for Java-Python Program Translation}

\author{
Wasi Uddin Ahmad$^\dagger$, Md Golam Rahman Tushar$^\S$ \\
{\bf Saikat Chakraborty}$^\ddagger$, {\bf Kai-Wei Chang}$^\dagger$ \\ [2pt]
$^\dagger$University of California, Los Angeles, $^\ddagger$Columbia University, $^\S$Independent Contributor \\ [2pt]
$^\dagger${\{wasiahmad, kwchang\}@cs.ucla.edu} \\ $^\ddagger${saikatc@cs.columbia.edu}, $^\S$grtushar11@gmail.com 
\\ [5pt]
{\url{https://github.com/wasiahmad/AVATAR}}
}


\begin{document}
\maketitle

\begin{abstract}
Program translation refers to migrating source code from one programming language (PL) to another. It has tremendous practical value in software development, as porting software across languages is time-consuming and costly. Automating program translation is of paramount importance in software migration, and recently researchers explored unsupervised approaches due to the unavailability of parallel corpora. However, the availability of pre-trained language models (PLMs) for PLs enables supervised fine-tuning with a small number of labeled examples. Therefore, we present \tool, a collection of 9,515 programming problems and their solutions written in two popular languages, Java and Python. \tool is collected from competitive programming sites, online platforms, and open-source repositories. Furthermore, \tool includes unit tests for 250 examples to facilitate functional correctness evaluation. We benchmark several PLMs fine-tuned on \tool. Experiment results show that the models lack in generating functionally accurate code.


\end{abstract}

\section{Introduction}

Software developers and researchers often require to convert software codebases or research prototypes from one platform to another or rewrite them in the target programming languages.
Manually rewriting software is time-consuming, expensive, and requires expertise in both the source and target languages. For example, the Commonwealth Bank of Australia spent around \$750 million and 5 years translating its platform from COBOL to Java \cite{lachaux2020unsupervised}.
Program translation system that converts the source code of a program written in a programming language to an equivalent program in a different programming language is known as transcompiler, transpiler, or source-to-source compiler.
Transcompilers have a prodigious practical value; they could help to reduce the translation efforts of developers and researchers by not requiring them to write code from scratch, instead they can edit the translated code with less efforts.

The conventional transcompilers are based on rule-based approaches; they first convert source code into an Abstract Syntax Tree (AST) and then apply handwritten rules to translate to the target language.
Development and adaptation of transcompilers need advanced knowledge and therefore are available in a handful of programming languages.
Undoubtedly, the automation of program translation would facilitate software development and research tremendously.

With the recent advancements in data-driven neural machine translation (NMT) approaches between natural languages, researchers have started investigating them for programming language translation.
\citet{lachaux2020unsupervised} trained an NMT system in an unsupervised fashion using large-scale monolingual source code from GitHub that showed noteworthy success in source code translation between Java, Python, and C++ languages.
Pre-trained language models (PLMs) of code
have been shown to work well on Java-C\# translation after fine-tuning on a small amount of parallel examples \cite{feng-etal-2020-codebert, guo2020graphcodebert, ahmad2021unified, wang-etal-2021-codet5}.
Motivated by these favourable results, in this work, we propose a new parallel corpus of Java and Python programs.

\begin{table*}[!ht]
\centering
\begin{tabular}{l|c|c c|c c|c|c|c}
\hline
\multirow{2}{*}{Source}  & 
\multirow{2}{*}{\#Prob.} & 
\multicolumn{2}{c|}{Java} & \multicolumn{2}{c|}{Python} & 
\multirow{2}{*}{Soln. / Prob.} & \multirow{2}{*}{Train} & \multirow{2}{*}{Valid / Test} \\
\cline{3-6}
& & \#Soln. & Avg\textsubscript{L} & \#Soln. & Avg\textsubscript{L} & & \\
\hline
AtCoder & 871 & 3,990 & 276.5 & 4,344 & 180.3 & [1 -- 5] & 14,604 & 36 / 195 \\
Code Jam & 120 & 508 & 390.9 & 460 & 266.5 & [1 -- 5] & 1,586/7 & 7/ 19 \\
Codeforces & 2,193 & 6,790 & 246.2 & 10,383 & 123.8 & [1 -- 5] & 24,754 & 102 / 436 \\
GeeksforGeeks & 5,019 & 5,019 & 194.8 & 5,019 & 138.4 & 1 & 3,754 & 269 / 996 \\
LeetCode & 107 & 107 & 140.0 & 107 & 97.4 & 1 & 82 & 7 / 18 \\
Project Euler & 162 & 162 & 227.3 & 162 & 139.4 & 1 & 110 & 11 / 41 \\
AIZU & 1,043 & 4,343 & 304.2 & 4,603 & 171.3 & [1 -- 5] & 15,248 & 44 / 199 \\ \hdashline
Total & 9,515 & 20,919 & 254.5 & 25,078 & 147.9 & - & 60,138 & 476 / 1,906 \\
\hline
\end{tabular}
\vspace{-2mm}
\caption{
Statistics of the \tool dataset. Avg\textsubscript{L} indicates the average program length (after parsing) written in Java and Python languages. We split the dataset into 75:5:20 to form training, validation and test examples, respectively. To form parallel examples for training, we pair up solutions in Java and Python, for validation and test examples, we consider multiple solutions as ground-truth.
}
\label{table:data_stat}
\vspace{-2.5mm}
\end{table*}

We propose a corpus, \tool (jAVA-pyThon progrAm tRanslation) that consists of solutions written in Java and Python for 9,515 programming problems collected from competitive programming sites, online platforms, and open source repositories.
\tool includes 250 examples with unit tests to facilitate functional correctness evaluation of program translation.
We train several baselines, including models trained from scratch or pre-trained on large-scale source code collection and fine-tuned on \tool.
The experiment results indicate that while the models perform
considerably in terms of the lexical match, they lack 
Furthermore, \tool offers 3,391 parallel functions that we use to train models or fine-tune pre-trained language models and perform function translation evaluation on the dataset released by  \citet{lachaux2020unsupervised}.



\section{\tool Construction}

\paragraph{Data Collection} 
We construct \tool based on solutions of computational problems written in Java and Python collected from open source programming contest sites: AtCoder, AIZU Online Judge, Google Code Jam, Codeforces, and online platforms: GeeksforGeeks, LeetCode, Project Euler.
We crawl Codeforces and GeeksforGeeks sites to collect the problem statements and their solutions.
We collect the AtCoder and AIZU data from \citet{puri2021codenet}, Google Code Jam data from \citet{nafi2019clcdsa}\footnote{\url{https://github.com/Kawser-nerd/CLCDSA}}, and LeetCode and Project Euler problem solutions from open source Github repositories.\footnote{\url{https://github.com/qiyuangong/leetcode}}\footnote{\url{https://github.com/nayuki/Project-Euler-solutions}}
We collect [1 -- 20] accepted solutions for a single problem written in Java and Python.

\paragraph{Preprocessing \& Filtering} 
At first, we tokenize the solution code and remove docstrings and comments from them.
We use the \texttt{javalang}\footnote{\url{https://github.com/c2nes/javalang}} tokenizer for Java and the \texttt{tokenizer}\footnote{\url{https://docs.python.org/3/library/tokenize.html}} of the standard library for Python.
After tokenization, we filter out solutions that are longer than a specified length threshold ($=$ 464).
In the initial data collection, there are [1 -- 20] accepted solutions for each problem.
We filter out solutions and only keep at most 5 solutions per problem.
Our goal is to keep the solutions that are maximal different from others in order to increase diversity among solutions of the same problem.
We use the open source library \texttt{difflib}\footnote{\url{https://docs.python.org/3/library/difflib.html}} to compare all the solutions pairwise (individually in Java and Python) and select five solutions that differ most from others.



\paragraph{Data Statistics}
We split 9,515 problem statements into 75:5:20 ratio to form 7,133 training, 476 validation, and 1,906 test examples.
Table \ref{table:data_stat} summarizes the data statistics.
Since we collect [1 -- 5] accepted solutions for each problem statement in both languages, we form [1 -- 25] parallel examples per problem for training. In evaluation, we use multiple ground-truths and select the best performance according to the evaluation metrics.

\paragraph{Unit Tests}
\tool presents unit tests for 250 evaluation examples (out of 1,906) to perform functional accuracy evaluation of the translation models.
The unit tests are collected from the publicly available test cases released by AtCoder.\footnote{\url{https://atcoder.jp/posts/21}}

\paragraph{Parallel Functions}
\tool includes 3,391 parallel Java and Python functions.\footnote{Deduplicated against the evaluation dataset released by \citet{lachaux2020unsupervised} using \url{https://github.com/microsoft/dpu-utils}.} 
The functions are extracted by parsing programs that include \emph{only} one function. We use them to train models and evaluate using the dataset released by \citet{lachaux2020unsupervised}.

\section{Experiment \& Results}

\begin{table*}[!ht]
\centering
\resizebox{\linewidth}{!}
{%
\begin{tabular}{l|c c c c c c|c c c c c c}
\hline
\multirow{2}{*}{{Models}} & \multicolumn{6}{c|}{{Java to Python}} & \multicolumn{6}{c}{{Python to Java}} \\ 
\cline{2-13}
& BLEU & SM & DM & CB & EA & CA & BLEU & SM & DM & CB & EA & CA \\ 
\hline
TransCoder & 38.7 & 31.6 & 38.2 & 36.4 & 77.3 & 0 & 45.2 & 39.3 & 20.1 & 32.4 & 0 & 0 \\
DOBF & 42.0 & 32.9 & {\bf 42.9} & 38.9 & 78.3 & 0 & 42.3 & 39.5 & 19.0 & 31.2 & 0 & 0 \\
TransCoder-ST & 41.7 & 33.1 & 42.6 & 39.3 & 85.8 & 0 & 42.5 & 37.4 & 20.4 & 30.7 & 0 & 0 \\ \hdashline
Seq2Seq+Attn. & 57.4 & 40.9 & 34.8 & 42.6 & 92.2 & 2.8 & 59.5 & 50.1 & 26.6 & 43.0 & 48.4 & 0.8 \\
Transformer & 39.6 & 35.0 & 33.5 & 34.8 & 92.3 & 0.4 & 43.5 & 44.9 & 25.2 & 35.6 & 63.8 & 0.4 \\ \hdashline
CodeGPT & 46.3 & 32.2 & 22.2 & 30.2 & 79.4 & 2.8 & 48.9 & 42.7 & 34.1 & 38.0 & 40.7 & {\bf 2.0} \\
CodeGPT-adapted & 44.3 & 31.6 & 20.4 & 29.3 & 80.2 & 2.4 & 48.0 & 43.0 & 28.3 & 36.7 & 46.7 & 0.8 \\
CodeBERT & 51.1 & 34.4 & 29.2 & 35.0 & 92.8 & 0.4 & 35.1 & 41.1 & 31.5 & 33.2 & 54.1 & 0 \\ 
GraphCodeBERT & 57.9 & 38.0 & 32.2 & 39.0 & 92.9 & 2.0 & 38.3 & 42.6 & 32.7 & 36.9 & 66.8 & 0 \\
PLBART & {\bf 63.1} & {\bf 42.2} & 37.9 & {\bf 46.2} & {\bf 96.4} & {\bf 6.8} & {\bf 69.7} & 54.2 & 30.9 & 48.8 & {\bf 78.3} & 0.8 \\ 
CodeT5 & 62.7 & 41.7 & 37.9 & {\bf 46.2} & 91.8 & 6.0 & 60.8 & {\bf 55.1} & {\bf 39.6} & {\bf 50.3} & 68.7 & 1.6 \\
TransCoder-ST & 55.4 & 41.6 & 36.1 & 43.8 & 94.9 & 5.6 & 66.0 & 53.3 & 31.7 & 48.6 & 72.4 & {\bf 2.0} \\
\hline
\end{tabular}
}
\vspace{-2mm}
\caption{
Test set results on our proposed corpus for Java-Python program translation. SM, DM, CB, EA, and CA stand for Syntax Match, Dataflow Match, CodeBLEU, Execution Accuracy, Computational Accuracy, respectively. 
}
\label{table:prog_trans}
\vspace{-2.5mm}
\end{table*}

\subsection{Evaluation Metrics}



\noindent\textbf{BLEU}
computes the overlap between candidate and reference translations \cite{papineni-etal-2002-bleu}. 


\noindent\textbf{Syntax Match (SM)} 
represents the percentage of the sub-trees extracted from the candidate program's abstract syntax tree (AST) that match the sub-trees in reference programs' AST.




\noindent\textbf{Dataflow Match (DM)} 
is the ratio of the number of matched candidate data-flows and the total number of the reference data-flows \cite{ren2020codebleu}.



\noindent\textbf{CodeBLEU (CB)} 
is the weighted avergae of token level match, syntax level match (SM), and Dataflow match (DM) \cite{ren2020codebleu}.



\noindent\textbf{Execution Accuracy (EA)}
indicates the percentage of translated programs that are executable (results in no compilation or runtime errors).

\noindent\textbf{Computational Accuracy (CA)} 
\citet{lachaux2020unsupervised} proposed the metric to evaluate whether the candidate translation generates the same outputs as the reference when given the same inputs.

\subsection{Models}
We evaluate a variety of models on program and function translation using \tool and the evaluation dataset released by \citet{lachaux2020unsupervised}.

\paragraph{Zero-shot}
This set of models are evaluated on \tool without any training or fine-tuning.

\noindent$\bullet$~\textbf{TransCoder} 
is pre-trained in an unsupervised fashion that can translate programs between Java, Python, and C++ languages \cite{lachaux2020unsupervised}.

\noindent$\bullet$~\textbf{DOBF} uses deobfuscation pretraining followed by unsupervised translation \cite{lachaux2021dobf}.

\noindent$\bullet$~\textbf{TransCoder-ST} is developed by fine-tuning TransCoder on a parallel corpus created via an automated unit-testing system \cite{roziere2021leveraging}.




\paragraph{Models trained from scratch}
These models are trained from scratch using \tool.
We use the sentencepiece tokenizer and vocabulary from \citet{ahmad2021unified} in these models.


\noindent$\bullet$~{\textbf{Seq2Seq+Attn.}} is an LSTM based sequence-to-sequence (Seq2Seq) model with attention mechanism \cite{bahdanau2014neural}. 

\noindent$\bullet$~{\textbf{Transformer}} 
is a self-attention based Seq2Seq model \cite{vaswani2017attention}. We use the Transformer architecture studied in \citet{ahmad-etal-2020-transformer}.

\paragraph{Pre-trained Models}
We evaluated three types of pre-trained models (PLMs). 
First, we evaluate decoder-only PLMs (\eg CodeGPT) that generate auto-regressively. 
The second category of PLMs is encoder-only (\eg CodeBERT). We use a randomly initialized decoder to finetune such models in a Seq2Seq fashion. 
The third category of PLMs is Seq2Seq models (\eg PLBART), which we directly finetune on translation tasks.


\noindent$\bullet$~\textbf{CodeGPT and CodeGPT-adapted} are GPT-2 \cite{radford2019language} style models pre-trained on CodeSearchNet~\cite{CodeXGLUE}. Note that CodeGPT-adapted starts from the GPT-2 checkpoint, while CodeGPT is pre-trained from scratch.


\noindent$\bullet$~\textbf{CodeBERT} is an encoder-only model that is pre-trained on unlabeled source code via masked language modeling (MLM) and replaced token detection objectives \cite{feng-etal-2020-codebert}.


\noindent$\bullet$~\textbf{GraphCodeBERT} is pre-trained using MLM, data flow edge prediction, and variable-alignment between code and its' data flow \cite{guo2020graphcodebert}.


\noindent$\bullet$~\textbf{PLBART} is a Transformer LM pre-trained via denoising autoencoding~\cite{ahmad2021unified}.

\noindent$\bullet$~\textbf{CodeT5} is a Transformer LM pre-trained via identifier-aware denoising \cite{wang-etal-2021-codet5}.

In addition, we fine-tune TransCoder-ST that is the best translation model in the literature.

\begin{table*}[!ht]
\centering
\resizebox{\linewidth}{!}
{%
\begin{tabular}{l|c c c c c c|c c c c c c}
\hline
\multirow{2}{*}{{Models}} & \multicolumn{6}{c|}{{Java to Python}} & \multicolumn{6}{c}{{Python to Java}} \\ 
\cline{2-13}
& BLEU & SM & DM & CB & EA & CA & BLEU & SM & DM & CB & EA & CA \\ 
\hline
TransCoder & 72.4 & 55.7 & 65.7 & 67.9 & 69.2 & 49.1 & 65.4 & 72.6 & 70.3 & 70.7 & 58.9 & 35.7 \\
DOBF & 72.2 & 56.6 & 63.7 & 67.5 & 73.1 & 52.2 & 67.7 & 72.8 & 69.4 & 71.2 & 63.5 & 44.4 \\
TransCoder-ST & 73.1 & 57.0 & {\bf 66.3} & 68.7 & 86.6 & 68.5 & 70.0 & 73.0 & 69.5 & 71.9 & 68.3 & 58.1 \\ \hdashline
Seq2Seq+Attn. & 50.9 & 53.6 & 55.2 & 56.6 & 51.5 & 28.9 & 29.5 & 44.0 & 13.5 & 29.3 & 18.0 & 1.5 \\
Transformer & 38.5 & 35.3 & 40.7 & 41.2 & 42.0 & 2.59 & 40.6 & 50.9 & 20.4 & 38.5 & 19.9 & 1.7 \\ \hdashline
CodeGPT & 64.9 & 53.2 & 52.7 & 59.3 & 65.9 & 41.8 & 49.2 & 54.9 & 48.5 & 51.3 & 47.3 & 31.1 \\
CodeGPT-adapted & 67.4 & 56.3 & 55.1 & 62.0 & 68.8 & 50.4 & 59.0 & 62.6 & 56.1 & 59.7 & 49.8 & 35.9 \\
CodeBERT & 52.0 & 45.6 & 41.5 & 48.9 & 45.5 & 10.4 & 45.4 & 54.9 & 32.6 & 45.0 & 25.7 & 4.2 \\ 
GraphCodeBERT & 58.6 & 49.6 & 46.9 & 54.5 & 46.8 & 18.3 & 51.9 & 58.9 & 37.4 & 50.4 & 27.0 & 10.0 \\
PLBART & {\bf 79.9} & {\bf 64.9} & 64.8 & {\bf 73.2} & {\bf 88.4} & 68.9 & 80.5 & {\bf 78.6} & 67.4 & 76.8 & 70.1 & 57.5 \\
CodeT5 & 79.4 & 64.1 & 63.2 & 72.5 & 83.8 & 61.0 & 79.0 & 77.1 & 67.7 & 75.9 & 64.3 & 52.7 \\
TransCoder-ST & 79.3 & 64.2 & 64.7 & 72.9 & 87.5 & {\bf 69.4} & {\bf 81.4} & {\bf 78.6} & {\bf 72.1} & {\bf 78.4} & {\bf 73.7} & {\bf 62.0} \\
\hline
\end{tabular}
}
\vspace{-2mm}
\caption{
Test set results on our proposed corpus for Java-Python function translation. SM, DM, CB, EA, and CA stand for Syntax Match, Dataflow Match, CodeBLEU, Execution Accuracy, Computational Accuracy, respectively.  
}
\label{table:fn_trans}
\vspace{-2.5mm}
\end{table*}

\subsection{Results}

\paragraph{Program Translation}
The performance comparison of all the experiment models is presented in Table \ref{table:prog_trans}.
In general all the experiment models perform well in terms of match-based metrics, \eg BLEU and CodeBLEU. However, the computational accuracy (CA) clearly indicates that these models are far from perfect in generating functionally accurate translations.
Overall, the best performing model is PLBART, resulting in highest execution accuracy (EA) and CA in Java to Python translation.
Note that, the zero EA score of TransCoder, DOBF, and TransCoder-ST in Python to Java translation is due to not generating a class correctly that fails execution of all translated programs.

\paragraph{Function Translation}
The performance comparison of all the experiment models is presented in Table \ref{table:fn_trans}. Apart from models trained from scratch, CodeBERT, and GraphCodeBERT, all the models perform well in terms of match-based metrics, execution and computational accuracy.
Overall, the best performing model is fine-tuned TransCoder-ST and PLBART is the closest competitor model.

We present execution evaluation breakdown and qualitative examples in Appendix \ref{appendix:anal} and \ref{appendix:qual_anal}.


\section{Related Works}

Several works in the past have contributed to build parallel corpus for source code translation.
\citet{nguyen2013lexical} curated the first parallel corpus of Java and C\# functions by developing a semi-automatic tool to search for the similar class names and method signatures from two open source projects, {\tt Lucene} and {\tt Db4o}.
Similarly, \citet{karaivanov2014phrase} built a mining tool that uses the Java and C\# ANTLR grammar to search for similar methods from five open source projects - {\tt Db4o}, {\tt Lucene}, {\tt Hibernate}, {\tt Quartz}, and {\tt Spring}.
Subsequent works used libraries and transcompilers to construct parallel corpus.
For example, \citet{aggarwal2015using} used {\tt 2to3}, a Python library\footnote{\url{https://docs.python.org/2/library/2to3}} 
and \citet{NEURIPS2018_d759175d} used a transcompiler to create a parallel corpus between Python 2 -- Python 3 and CoffeeScript -- Javascript, respectively.
Recently, \citet{lachaux2020unsupervised} collected programming problem solutions in Java, Python, and C++ ($\sim$850 functions in each language) from {\tt GeeksforGeeks} to evaluate their proposed translation model.
Concurrent works \citep{CodeGeeX, athiwaratkun2022multi} present unit tests based benchmark to evaluate zero-shot translation capabilities of large language models.
Different from the these works, we propose a sizeable parallel corpus of Java and Python programs by collecting programming problem solutions from competitive programming sites, online platforms, and open source repositories.


\section{Conclusion}
This work proposes a parallel corpus of Java and Python programs to contribute to the development of translation systems for programming languages that have a sizeable impact in software development. We evaluate several neural machine translation systems on the proposed dataset and perform analysis to reveal crucial factors that affect program translation accuracy. In our future work, we want to increase the size of the parallel corpus and support more programming languages.

\clearpage
\section*{Limitations}
The proposed benchmark has a few limitations. First, \tool has a smaller training data size which limits training deep neural models from scratch. Second, the dataset covers only two programming languages. Third, \tool includes parallel examples of programs and functions that mostly focus on the use of data structures and algorithms. On the other hand, most software developers write program as part of software projects that include API dependencies. Therefore, it is unknown whether \tool could facilitate program or function translation for such settings. Due to a lack of computational resources, we could not evaluate large language models (LLMs) \cite{nijkamp2022conversational, fried2022incoder, CodeGeeX}. Therefore, it is unknown how much \tool could bring value for LLMs. However, our code release would help evaluating LLMs.

\section*{Ethics Statement}

\paragraph{License} 
The \texttt{LeetCode} examples we crawled from the Github repository is under MIT license. On the other hand, \texttt{Project Euler} and \texttt{Code Jam} examples collected from Github does not have any license information. The \texttt{AtCoder} and \texttt{AIZU} examples are collected from \texttt{CodeNet} which is under Apache-2.0 license. We crawl examples from \texttt{GeeksforGeeks} and \texttt{Codeforces} and release them under CC BY-NC-SA 4.0 license. To use the \tool benchmark, we are required to strictly adhere to these licenses.

\paragraph{Carbon Footprint} We avoided fine-tuning large models due to computational limitation, resulting in reduced impact to the environment. We finetuned nine models on program and function translation tasks and due to the smaller size of the training data, all jobs took a total of 1--2 days on \textsf{RTX 2080 Ti} GPUs. A total 100 hours of training in a single \textsf{RTX 2080 Ti} GPU results in approximately 7.5kg of carbon emission into the environment.\footnote{Estimations were conducted using the \href{https://mlco2.github.io/impact\#compute}{MachineLearning Impact calculator} presented in \cite{lacoste2019quantifying}. We use Amazon Web Services as the provider.}

\paragraph{Sensitive Information} \tool composed of parallel programs and functions that do not have any natural language (NL) comments or docstring. We remove them to get rid of any personally identifiable information or offensive content. However, there could still be such content in the form of \emph{string} as we do not manually check each example.

%

\bibliography{anthology,custom}
\bibliographystyle{acl_natbib}

\clearpage
\appendix

\twocolumn[{%
 \centering
 \Large\bf Supplementary Material: Appendices \\ [20pt]
}]

\section{Hyperparameters Details}
We individually fine-tune the models for Java to Python and Python to Java program and function translation, respectively.
We fine-tune the models for a maximum of 20 epochs using the Adam~\citep{kingma2015adam} optimizer with a batch size of 32. 
We tune the learning rate in the range $[1e-4, 5e-5, 3e-5, 1e-5]$.
The final models are selected based on the validation BLEU score.
We use beam decoding with beam size set to 10 for inference across all the models.

\section{Analysis}
\label{appendix:anal}

\paragraph{Execution-based Evaluation Breakdown}
We present the breakdown for the testcase based evaluation in Table \ref{table:error_breakdown}.
For program translation evaluation, \tool consists of 250 evaluation examples with unit tests. For function translation evaluation, we use the test examples released by \cite{lachaux2020unsupervised}. Among the examples, 464 Java to Python and 482 Python to Java examples have test cases. We present the number of success, failure, and error counts in Table \ref{table:error_breakdown}.
We further present the compilation and runtime error breakdown in Table \ref{table:ce_vs_re}.

To analyze program translation errors, we manually examine the errors made by PLBART. We observe that PLBART does not generate the import statements in Java properly, resulting in many failures to find symbols (\eg StringTokenizer, BufferedReader). Moreover, a quick look at the error made by all models reveals that \emph{type mismatch} is one of the primary causes of compilation errors in all the models. We also notice that models fail to translate longer programs.

\begin{table*}[!ht]
\centering
\resizebox{\linewidth}{!}
{%
\begin{tabular}{l@{\hskip 0.1in} | c@{\hskip 0.1in} c@{\hskip 0.1in} c@{\hskip 0.1in} c@{\hskip 0.1in} c@{\hskip 0.05in} | c@{\hskip 0.1in} c@{\hskip 0.1in} c@{\hskip 0.1in} c@{\hskip 0.1in} c@{\hskip 0.05in}}
\hline
\multirow{2}{*}{{Models}} & \multicolumn{5}{c|}{{Java to Python}} & \multicolumn{5}{c}{{Python to Java}} \\ 
\cline{2-11}
& \#Tests & Error & Failure & Timeout & Success & \#Tests & Error & Failure & Timeout & Success \\ 
\hline
\multicolumn{11}{l}{{\bf Program Translation}} \\
\hline
TransCoder          & 250 & 53 & 197 & 0 & 0 & 250 & 250 & 0 & 0 & 0 \\
DOBF                & 250 & 62 & 188 & 0 & 0 & 250 & 250 & 0 & 0 & 0 \\
TransCoder-ST       & 250 & 55 & 195 & 0 & 0 & 250 & 250 & 0 & 0 & 0 \\ \hdashline
Seq2Seq+Attn.       & 250 & 143 & 98 & 2 & 7 & 250 & 218 & 30 & 0 & 2 \\
Transformer         & 250 & 156 & 92 & 1 & 1 & 250 & 246 & 3 & 0 & 1 \\ \hdashline
CodeGPT             & 250 & 140 & 102 & 1 & 7 & 250 & 169 & 76 & 0 & 5 \\
CodeGPT-adapted     & 250 & 119 & 121 & 4 & 6 & 250 & 245 & 3 & 0 & 2 \\
CodeBERT            & 250 & 189 & 57 & 3 & 1 & 250 & 248 & 2 & 0 & 0 \\
GraphCodeBERT       & 250 & 93 & 147 & 5 & 5 & 250 & 216 & 34 & 0 & 0 \\
PLBART              & 250 & 102 & 124 & 7 & 17 & 250 & 241 & 6 & 1 & 2 \\
CodeT5              & 250 & 111 & 119 & 5 & 15 & 250 & 226 & 20 & 0 & 4 \\
TransCoder-ST       & 250 & 135 & 92 & 9 & 14 & 250 & 194 & 51 & 0 & 5 \\
\hline
\multicolumn{11}{l}{{\bf Function Translation}} \\
\hline
TransCoder          & 464 & 143 & 89 & 4 & 228 & 482 & 198 & 106 & 6 & 172 \\
DOBF                & 464 & 125 & 88 & 9 & 242 & 482 & 176 & 88 & 4 & 214 \\
TransCoder-ST       & 464 & 62 & 79 & 5 & 318 & 482 & 153 & 48 & 1 & 280 \\ \hdashline
Seq2Seq+Attn.       & 464 & 225 & 97 & 8 & 134 & 482 & 395 & 77 & 3 & 7 \\
Transformer         & 464 & 269 & 170 & 13 & 12 & 482 & 386 & 83 & 5 & 8 \\ \hdashline
CodeGPT             & 464 & 158 & 103 & 9 & 194 & 482 & 254 & 74 & 4 & 150 \\
CodeGPT-adapted     & 464 & 145 & 78 & 7 & 234 & 482 & 242 & 64 & 3 & 173 \\
CodeBERT            & 464 & 253 & 149 & 14 & 48 & 482 & 358 & 94 & 10 & 20 \\
GraphCodeBERT       & 464 & 247 & 118 & 14 & 85 & 482 & 352 & 80 & 2 & 48 \\
PLBART              & 464 & 54 & 91 & 4 & 315 & 482 & 144 & 58 & 3 & 277 \\
CodeT5              & 464 & 75 & 97 & 9 & 283 & 482 & 172 & 51 & 5 & 254 \\
TransCoder-ST       & 464 & 58 & 79 & 5 & 322 & 482 & 127 & 51 & 5 & 299 \\
\hline
\end{tabular}
}
\caption{
Breakdown of the success, error, failure, and timeout in the execution based evaluation. \#Tests indicates the number of evaluation examples with unit tests. While success indicates the number of examples passing all the unit tests, Failure indicates number of examples that did not at least one of the unit tests. The Error count indicates number of examples with compilation and runtime errors.
}
\label{table:error_breakdown}
\end{table*}

\begin{table*}[!ht]
\centering
{%
\begin{tabular}{l|c c c| c c c}
\hline
\multirow{2}{*}{{Models}} & \multicolumn{3}{c|}{{Java to Python}} & \multicolumn{3}{c}{{Python to Java}} \\ 
\cline{2-7}
& \#Tests & CE & RE & \#Tests & CE & RE \\ \hline
TransCoder          & 464 & 0\% & 30.8\% & 482 & 31.3\% & 9.8\% \\
DOBF                & 464 & 0\% & 26.9\% & 482 & 27.4\% & 9.1\% \\
TransCoder-ST       & 464 & 0\% & 13.4\% & 482 & 24.9\% & 6.9\% \\ 
\hdashline
Seq2Seq+Attn.       & 464 & 0\% & 48.5\% & 482 & 80.3\% & 1.5\% \\
Transformer         & 464 & 0\% & 58.0\% & 482 & 78.0\% & 2.1\% \\ 
\hdashline
CodeGPT             & 464 & 0\% & 34.1\% & 482 & 49.4\% & 2.7\% \\
CodeGPT-adapted     & 464 & 0\% & 31.3\% & 482 & 46.7\% & 3.3\% \\
CodeBERT            & 464 & 0\% & 54.5\% & 482 & 71.4\% & 2.7\% \\
GraphCodeBERT       & 464 & 0\% & 53.2\% & 482 & 71.8\% & 1.2\% \\
PLBART              & 464 & 0\% & 11.6\% & 482 & 25.3\% & 4.6\% \\
CodeT5              & 464 & 0\% & 16.2\% & 482 & 32.4\% & 3.1\% \\
TransCoder-ST       & 464 & 0\% & 12.5\% & 482 & 22.6\% & 3.7\% \\
\hline
\end{tabular}
}
\caption{
Compilation error (CE) vs. runtime error (RE) percentage in function translation. 
}
\label{table:ce_vs_re}
\end{table*}

\section{Qualitative Examples}
\label{appendix:qual_anal}

\paragraph{Program Translation}
We demonstrate a couple of qualitative examples of Java to Python program translation by PLBART in Figure \ref{tab:qual_ex1}.
We observe that PLBART correctly translates Java API \texttt{Math.pow()} to \texttt{pow()} in Python. We also observe that PLBART learns to translate a class with a function in Java to a function only in Python. 

In Figure \ref{tab:qual_ex2}, we present an example of Python to Java program translation. We see PLBART fail to translate correctly. We notice PLBART unnecessarily generates \texttt{InputReader} class that uses \texttt{BufferedReader} to read from standard input.
Furthermore, we observed another behavior that when translating from Python to Java, PLBART generates class with name either \texttt{Main} or \texttt{GFG}. This is presumably due to the generic class name used in many programming solutions and \texttt{GeeksforGeeks} examples.

\paragraph{Function Translation}
We present qualitative examples of Java to Python and Python to Java function translation by PLBART in Figure \ref{tab:qual_ex3} and \ref{tab:qual_ex4}. Overall, we observe pretty good quality of translations, although there are translations that do not pass all the unit tests as demonstrated by the computational accuracy in the main result.

\clearpage
\begin{figure*}[!htb]

\textsc{\underline{Input Program in Java}} \\
\begin{adjustbox}{valign=t,minipage=1.0\textwidth}
\begin{tabular}{l}
\lstset{escapechar=~,style=CustomJava}
\begin{lstlisting}
class GFG {
    static int cntSquares ( int n ) {
        int squares = 0 ;
        for ( int i = 1 ; i <= n ; i ++ ) {
            squares += Math . pow ( i , 2 ) ;
        }
        return squares ;
    }
    public static void main ( String args [ ] ) {
        int n = 4 ;
        System . out . print ( cntSquares ( 4 ) ) ;
    }
}
\end{lstlisting}
\end{tabular}
\end{adjustbox}

\medskip
\textsc{\underline{Java to Python Translation}} \\
\begin{adjustbox}{valign=t,minipage=1.0\textwidth}
\begin{tabular}{l}
\lstset{escapechar=@,style=CustomPy}
\begin{lstlisting}
def cntSquares ( n ) :
    squares = 0
    for i in range ( 1 , n + 1 ) :
        squares += pow ( i , 2 )
    return squares

n = 4
print ( cntSquares ( 4 ) )
\end{lstlisting}
\end{tabular}
\end{adjustbox}

\bigskip
\bigskip

\textsc{\underline{Input Program in Java}} \\
\begin{adjustbox}{valign=t,minipage=1.0\textwidth}
\begin{tabular}{l}
\lstset{escapechar=~,style=CustomJava}
\begin{lstlisting}
import java . util . Scanner ;

public class Main {
    public static void main ( String args [ ] ) throws Exception {
        Scanner sc = new Scanner ( System . in ) ;
        String s = sc . next ( ) ;
        int L = s . length ( ) ;
        int same = 0 ;
        if ( s . charAt ( 0 ) == s . charAt ( L - 1 ) ) {
            same = 1 ;
        }
        if ( ( L + same ) % 2 == 0 ) {
            System . out . println ( "Second" ) ;
        }
        else {
            System . out . println ( "First" ) ;
        }
    }
}
\end{lstlisting}
\end{tabular}
\end{adjustbox}

\medskip
\textsc{\underline{Java to Python Translation}} \\
\begin{adjustbox}{valign=t,minipage=1.0\textwidth}
\begin{tabular}{l}
\lstset{escapechar=@,style=CustomPy}
\begin{lstlisting}
s = input ( )
if s [ 0 ] == s [ - 1 ] :
    print ( "First" )
elif s [ 0 ] == s [ - 1 ] :
    print ( "Second" )
else :
    print ( "First" )
\end{lstlisting}
\end{tabular}
\end{adjustbox}

\caption{
Qualitative examples of program translation from Java to Python by PLBART.
}
\label{tab:qual_ex1}
\end{figure*}

\begin{figure*}[t]

\textsc{\underline{Input Program in Python}} \\
\begin{adjustbox}{valign=t,minipage=1.0\textwidth}
\begin{tabular}{l}
\lstset{escapechar=~,style=CustomJava}
\begin{lstlisting}
S = input ( )
if S [ 0 ] == S [ - 1 ] :
    if len ( S ) % 2 == 0 :
        print ( "First" )
    else :
        print ( "Second" )
else :
    if len ( S ) % 2 == 1 :
        print ( "First" )
    else :
        print ( "Second" )
\end{lstlisting}
\end{tabular}
\end{adjustbox}

\medskip
\textsc{\underline{Python to Java Translation}} \\
\begin{adjustbox}{valign=t,minipage=1.0\textwidth}
\begin{tabular}{l}
\lstset{escapechar=@,style=CustomPy}
\begin{lstlisting}
import java . io . OutputStream ;
import java . io . IOException ;
import java . io . InputStream ;
import java . io . PrintWriter ;
import java . util . InputMismatchException ;
import java . io . IOException ;
import java . io . InputStream ;
import java . io . InputStream ;

public class Main {
    public static void main ( String [ ] args ) {
        InputStream inputStream = System . in ;
        OutputStream outputStream = System . out ;
        InputReader in = new InputReader ( inputStream ) ;
        PrintWriter out = new PrintWriter ( outputStream ) ;
        Task solver = new Task ( ) ;
        solver . solve ( 1 , in , out ) ;
        out . close ( ) ;
    }
    static class Task {
        public void solve ( int testNumber , InputReader in , PrintWriter out ) {
            String s = in . next ( ) ;
            if ( s . length ( ) % 2 == 0 ) {
                out . println ( "First" ) ;
            }
            else {
                out . println ( "Second" ) ;
            }
        }
    }
    static class InputReader {
        public BufferedReader reader ;
        public StringTokenizer tokenizer ;
        public InputReader ( InputStream stream ) {
            reader = new BufferedReader ( new InputStreamReader ( stream ) , 32768 ) ;
            tokenizer = null ;
        }
        public String next ( ) {
            while ( tokenizer == null || ! tokenizer . hasMoreTokens ( ) ) {
                try {
                    tokenizer = new StringTokenizer ( reader . readLine ( ) ) ;
                }
                catch ( IOException e ) {
                    throw new InputMismatchException ( ) ;
                }
            }
            return tokenizer . nextToken ( ) ;
        }
        public int nextInt ( ) {
            return Integer . parseInt ( next ( ) ) ;
        }
    }
}
\end{lstlisting}
\end{tabular}
\end{adjustbox}

\caption{
A qualitative examples of program translation from Python to Java by PLBART.
}
\label{tab:qual_ex2}
\end{figure*}

\begin{figure*}[!htb]

\textsc{\underline{Input Function in Java}} \\
\begin{adjustbox}{valign=t,minipage=1.0\textwidth}
\begin{tabular}{l}
\lstset{escapechar=~,style=CustomJava}
\begin{lstlisting}
static int smallest ( int x , int y , int z ) { 
    if ( ( y / x ) != 1 ) return ( ( y / z ) != 1 ) ? y : z ; 
    return ( ( x / z ) != 1 ) ? x : z ; 
}
\end{lstlisting}
\end{tabular}
\end{adjustbox}

\medskip
\textsc{\underline{Translation in Python by PLBART}} \\
\begin{adjustbox}{valign=t,minipage=1.0\textwidth}
\begin{tabular}{l}
\lstset{escapechar=~,style=CustomPy}
\begin{lstlisting}
def smallest ( x , y , z ) : 
    if ( ( y // x ) != 1 ) : 
        return y 
    if ( ( y // z ) != 1 ) : 
        return x 
    return z
\end{lstlisting}
\end{tabular}
\end{adjustbox}

\bigskip
\bigskip

\textsc{\underline{Input Function in Java}} \\
\begin{adjustbox}{valign=t,minipage=1.0\textwidth}
\begin{tabular}{l}
\lstset{escapechar=~,style=CustomJava}
\begin{lstlisting}
static int maxvolume ( int s ) { 
    int length = s / 3 ; s -= length ; 
    int breadth = s / 2 ; 
    int height = s - breadth ; 
    return length * breadth * height ; 
}
\end{lstlisting}
\end{tabular}
\end{adjustbox}

\medskip
\textsc{\underline{Translation in Python by PLBART}} \\
\begin{adjustbox}{valign=t,minipage=1.0\textwidth}
\begin{tabular}{l}
\lstset{escapechar=~,style=CustomPy}
\begin{lstlisting}
def maxvolume ( s ) : 
    length = s // 3 
    s -= length
    breadth = s // 2
    height = s - breadth
    return length * breadth * height
\end{lstlisting}
\end{tabular}
\end{adjustbox}

\bigskip
\bigskip

\textsc{\underline{Input Function in Java}} \\
\begin{adjustbox}{valign=t,minipage=1.0\textwidth}
\begin{tabular}{l}
\lstset{escapechar=~,style=CustomJava}
\begin{lstlisting}
static void printClosest ( int arr [ ] , int n , int x ) { 
    int res_l = 0 , res_r = 0 ; 
    int l = 0 , r = n - 1 , diff = Integer . MAX_VALUE ; 
    while ( r > l ) { 
        if ( Math . abs ( arr [ l ] + arr [ r ] - x ) < diff ) { 
            res_l = l ; 
            res_r = r ; 
            diff = Math . abs ( arr [ l ] + arr [ r ] - x ) ; 
        } 
        if ( arr [ l ] + arr [ r ] > x ) r -- ; 
        else l ++ ; 
    } 
    System . out . println ( " The closest pair is " + arr [ res_l ] + " and " + arr [ res_r ] ) ; 
}
\end{lstlisting}
\end{tabular}
\end{adjustbox}

\medskip
\textsc{\underline{Translation in Python by PLBART}} \\
\begin{adjustbox}{valign=t,minipage=1.0\textwidth}
\begin{tabular}{l}
\lstset{escapechar=~,style=CustomPy}
\begin{lstlisting}
def printClosest ( arr , n , x ) : 
    res_l = 0 
    res_r = 0 
    l = 0 
    r = n - 1 
    diff = sys . maxsize NEW_LINE while ( r > l ) : 
        if ( abs ( arr [ l ] + arr [ r ] - x ) < diff ) : 
            res_l = l 
            res_r = r 
            diff = abs ( arr [ l ] + arr [ r ] - x ) 
        if ( arr [ l ] + arr [ r ] > x ) : 
            r -= 1 
        else : 
            l += 1 
    print ( " The closest pair is " , arr [ res_l ] , " and " , arr [ res_r ] )
\end{lstlisting}
\end{tabular}
\end{adjustbox}

\caption{
Qualitative examples of function translation from Java to Python by PLBART.
}
\label{tab:qual_ex3}
\end{figure*}

\begin{figure*}[!htb]

\textsc{\underline{Input Function in Python}} \\
\begin{adjustbox}{valign=t,minipage=1.0\textwidth}
\begin{tabular}{l}
\lstset{escapechar=~,style=CustomPy}
\begin{lstlisting}
def minimum_cost ( a , n ) : 
    mn = sys . maxsize 
    sum = 0 
    for i in range ( n ) : 
        mn = min ( a [ i ] , mn ) 
        sum += a [ i ] 
    return mn * ( sum - mn )
\end{lstlisting}
\end{tabular}
\end{adjustbox}

\medskip
\textsc{\underline{Translation in Java by PLBART}} \\
\begin{adjustbox}{valign=t,minipage=1.0\textwidth}
\begin{tabular}{l}
\lstset{escapechar=~,style=CustomJava}
\begin{lstlisting}
static int minimum_cost ( int a [ ] , int n ) { 
    int mn = Integer . MAX_VALUE ; 
    int sum = 0 ; 
    for ( int i = 0 ; i < n ; i ++ ) { 
        mn = Math . min ( a [ i ] , mn ) ; 
        sum += a [ i ] ; 
    } 
    return mn * ( sum - mn ) ; 
}
\end{lstlisting}
\end{tabular}
\end{adjustbox}

\bigskip
\bigskip

\textsc{\underline{Input Function in Python}} \\
\begin{adjustbox}{valign=t,minipage=1.0\textwidth}
\begin{tabular}{l}
\lstset{escapechar=~,style=CustomPy}
\begin{lstlisting}
def nextPowerOf2 ( n ) : 
    p = 1 
    if ( n and not ( n & ( n - 1 ) ) ) : 
        return n 
    while ( p < n ) : 
        p <<= 1 
    return p
\end{lstlisting}
\end{tabular}
\end{adjustbox}

\medskip
\textsc{\underline{Translation in Java by PLBART}} \\
\begin{adjustbox}{valign=t,minipage=1.0\textwidth}
\begin{tabular}{l}
\lstset{escapechar=~,style=CustomJava}
\begin{lstlisting}
static int nextPowerOf2 ( int n ) { 
    int p = 1 ; 
    if ( n && ! ( n & ( n - 1 ) ) ) 
        return n ; 
    while ( p < n ) 
        p <<= 1 ; 
    return p ; 
}
\end{lstlisting}
\end{tabular}
\end{adjustbox}

\bigskip
\bigskip

\textsc{\underline{Input Function in Python}} \\
\begin{adjustbox}{valign=t,minipage=1.0\textwidth}
\begin{tabular}{l}
\lstset{escapechar=~,style=CustomPy}
\begin{lstlisting}
def isPerfectSquare ( n ) : 
    i = 1 
    the_sum = 0 
    while the_sum < n : 
        the_sum += i 
        if the_sum == n : 
            return True 
        i += 2 
    return False
\end{lstlisting}
\end{tabular}
\end{adjustbox}

\medskip
\textsc{\underline{Translation in Java by PLBART}} \\
\begin{adjustbox}{valign=t,minipage=1.0\textwidth}
\begin{tabular}{l}
\lstset{escapechar=~,style=CustomJava}
\begin{lstlisting}
static boolean isPerfectSquare ( int n ) { 
    int i = 1 , the_sum = 0 ; 
    while ( the_sum < n ) { 
        the_sum += i ; 
        if ( the_sum == n ) 
            return true ; 
        i += 2 ; 
    } 
    return false ; 
}
\end{lstlisting}
\end{tabular}
\end{adjustbox}

\caption{
Qualitative examples of function translation from Python to Java by PLBART.
}
\label{tab:qual_ex4}
\end{figure*}

\end{document}



\onecolumn
\begin{center}
    {\Large\bf Supplementary Materials}
\end{center}

\input{paper/appendix}

